\documentclass[lettersize,journal]{IEEEtran}
\usepackage{amsmath,amsfonts}
\usepackage{algorithmic}
\usepackage{algorithm}
\usepackage{array}
\usepackage{textcomp}
\usepackage{stfloats}
\usepackage{url}
\usepackage{verbatim}
\usepackage{graphicx}
\usepackage{xcolor}
\usepackage{cite}
\usepackage{subfigure}
\usepackage{cuted}
\newtheorem{theorem}{Theorem}
\newtheorem{lemma}{Lemma}
\IEEEaftertitletext{\vspace{-2.5\baselineskip}}
\begin{document}

\title{Efficient Design of Flag Sequence Sets for Delay-Doppler Estimation via Riemannian Optimization}

\author{Xuyang Zhao,~\IEEEmembership{Graduate Student Member,~IEEE,}
{Zeping Sui},~\IEEEmembership{Member,~IEEE,}
{Zilong Liu},~\IEEEmembership{Senior Member,~IEEE,}
{Jiangtao Wang},~\IEEEmembership{Member,~IEEE,}
and {Yongchao Wang},~\IEEEmembership{Senior Member,~IEEE}\thanks{Xuyang Zhao is with the School of Telecommunication Engineering, Xidian University, Xi'an 710071, China (e-mail: xy\_zhao@stu.xidian.edu.cn).

Zeping Sui and Zilong Liu are with the School of Computer Science and Electronic Engineering, University of Essex, CO4 3SQ Colchester, U.K. (e-mails: zepingsui@outlook.com; zilong.liu@essex.ac.uk)

Jiangtao Wang is with the Guangzhou Institute of Technology, Xidian University, Guangzhou, 510555, China (e-mail: jtwang@xidian.edu.cn)

Yongchao Wang is with the State Key Laboratory of Integrated Services Networks, Xidian University, Xi'an 710071, China, and also with the Guangzhou Institute of Technology, Xidian University, Guangzhou, 510555, China (e-mail: ychwang@mail.xidian.edu.cn)}}

\maketitle

\begin{abstract}
This letter investigates the design of a Flag sequence set with peak-curtain auto-ambiguity functions to enable low-complexity delay-Doppler estimation. Given the predefined Curtain sequences, the Flag sequence set design problem is formulated as the minimization of the weighted integrated masked sidelobe level. To solve this problem, we first reformulate it as an unconstrained optimization problem on the unimodular manifold. Then, a tailored unimodular manifold conjugate gradient (UM-CG) algorithm is developed. Compared with the existing define algorithm, the proposed UM-CG algorithm reduces computational and storage requirements. Moreover, the global convergence of the proposed UM-CG scheme is theoretically derived under mild conditions. Numerical results demonstrate the superior optimization performance of the proposed UM-CG algorithm and show that the designed Flag sequences exhibit nearly ideal peak-curtain~properties.
\end{abstract}


\vspace{-0.3cm}
\section{Introduction}
\vspace{-0.1cm}
High mobility induces time-varying channels and pronounced Doppler effects, posing significant challenges to communication and sensing systems and highlighting the importance of accurate delay-Doppler estimation\cite{Sui_ISAC}. In wireless communications, delay-Doppler estimation plays a key role in channel estimation and subsequent data detection\cite{OTFS_Survey}, whereas in radar sensing, delay and Doppler provide information on target range and radial velocity, respectively, forming the basis for target detection, localization, and tracking\cite{Radar_Signal}.

Delay-Doppler estimation typically relies on sequences with zero- or low-ambiguity zone (ZAZ/LAZ) properties \cite{ZAZ_LAZ}. However, achieving ideal auto-ambiguity function (AAF) properties over the entire delay-Doppler plane is challenging\cite{ideal_AAF}. Therefore, practical sequence designs focus on suppressing AAF sidelobes within a prescribed zone of operation (ZoO)\cite{ZoO}, while also reducing cross-ambiguity function (CAF) sidelobes between distinct sequences to mitigate mutual interference in multiple-input multiple-output and multiuser communication systems\cite{CAF}. Nevertheless, conventional delay-Doppler estimation requires an exhaustive two-dimensional search over the ZoO, resulting in excessive computational complexity that increases with the sequence length and the size\cite{delay_doppler_estimation_fast}. \looseness=-1\par

To this end, the concept of Flag sequence with a peak-curtain ambiguity function structure was proposed in \cite{Linear_complexity}, together with a systematic construction based on Heisenberg-Weil sequences, enabling near-linear-complexity estimation. However, these sequences suffer from high AAF sidelobes and prime-length restrictions. To address these limitations, arbitrary-length Curtain sequences and an accelerated parallel majorization-minimization (AP-MM) algorithm for Flag sequence design were introduced in \cite{Flag_sequence}. Nevertheless, AP-MM incurs substantial computational and memory overhead, while its squared iterative method acceleration lacks theoretical convergence guarantees and requires multiple variable updates per iteration \cite{MM_Song}.

As a parallel development, recent study of \cite{unimodular} has shown that manifold optimization provides an efficient and theoretically grounded approach for solving unimodular-constrained optimization problems. Motivated by this, we develop a tailored unimodular manifold conjugate gradient (UM-CG) algorithm to solve the optimization problem formulated in \cite{Flag_sequence}, where the transmit and receive reference Peak sequences are identical. Our main contributions are as follows: 1) We propose a computationally efficient UM-CG algorithm that directly tackles the original optimization problem without relaxation. Compared with AP-MM, the proposed algorithm requires only a single variable update per iteration and avoids the large-scale matrix expansion associated with MM based reformulation, thereby reducing both computational and storage overhead. 2) Unlike AP-MM, for which no theoretical convergence guarantee has been established, we derive the global convergence of the proposed UM-CG algorithm under mild conditions. Numerical simulations further demonstrate its effectiveness and computational efficiency. \looseness=-1\par

{\it Notations:} scalar variables are represented by standard lowercase letters, while vectors and matrices are denoted by lowercase and uppercase boldface letters, respectively. $\mathbb{C}^N$ and $\mathbb{C}^{N \times M}$ denote the sets of $N$-dimensional complex vectors and $N \times M$ complex matrices, respectively. The transpose, conjugate transpose and conjugate of a matrix $\mathbf{X}$ are represented as $\mathbf{X}^T$, $\mathbf{X}^{\dagger}$ and $\mathbf{X}^*$, respectively. The real-part operator is denoted by $\operatorname{Re}(\cdot)$, while $\circ$ represents the Hadamard product.  Furthermore, $\langle\cdot,\cdot\rangle_{\mathbf{X}}$ and $\|\cdot\|_{\mathbf{X}}$ are the inner product and its induced norm in the tangent space at $\mathbf{X}$, respectively. Finally, $\mathbf{0}_{N\times M}$ denotes the $N\times M$ zero matrix.
\vspace{-0.25cm}
\section{Preliminaries} \label{prelinminaries}
\vspace{-0.1cm}
We consider a delay-Doppler channel with $L$ propagation paths. When a length-$N$ time-domain sequence $\mathbf{s} \in \mathbb{C}^N$ is transmitted, the received signal can be expressed as
\begin{equation}
r[n]=\sum_{l=1}^{L}h_ls[n-\tau_l]e^{j\frac{2\pi}{N}\omega_l n}+w[n],
\end{equation}
for $n = 0, \cdots, N-1$, where $h_l$, $\tau_l$, and $\omega_l$ denote the complex-valued path gain, integer delay, and normalized Doppler frequency shift of the $l$-th path, respectively, while $w[n]$ represents additive white Gaussian noise. The delay-Doppler pairs satisfy $(\tau_l, \omega_l) \in \boldsymbol{\Gamma}$, where the ZoO $\boldsymbol{\Gamma}$ is specified as a discrete delay-Doppler grid with approximately $K$ points along each dimension, defined by $\boldsymbol{\Gamma}=\left\{(\tau, \omega) \mid \tau \in\left[-\tau_{\max }, \tau_{\max }\right], \omega \in\left[-\omega_{\max }, \omega_{\max }\right]\right\}$ with maximum delay $\tau_{\max}$ and maximum normalized Doppler frequency shift $\omega_{\max}$. \looseness=-1\par

To estimate the delay-Doppler pairs $\{(\tau_l,\omega_l)\}_{l = 1}^L$ associated with the propagation paths, a classical approach is the matched filter algorithm\cite{Linear_complexity}, yielding the $(\tau, \omega)$-th entry of matched filter matrix as
 \begin{equation}
     \mathbf{MF}[\tau, \omega] = \sum_{n=0}^{N-1} r[n]^*e^{j\frac{2\pi}{N}\omega n}s[n-\tau],
 \end{equation}
which corresponds to the ambiguity function (AF) between the transmit sequence and receive reference sequences. In this context, low-complexity delay-Doppler estimation can be achieved by exploiting the distinctive peak-curtain structure of Flag sequences. Specifically, for a Flag sequence with unit energy, its ideal AAF exhibits a peak with unit amplitude at the origin and a curtain with an amplitude of 0.5. \looseness=-1\par

Let $\mathbf{F}^s = \{\mathbf{f}_m^s\}_{m=1}^M$ and $\mathbf{F}^r = \{\mathbf{f}_m^r\}_{m=1}^M$ denote the transmit and receive reference Flag sequence sets, respectively, each consisting of $M$ sequences of length $N$. The $m$-th transmit Flag sequence and its corresponding receive reference Flag sequence are constructed as $\mathbf{f}_m^s = (\mathbf{c}_m^s + \mathbf{p}_m^s)/\sqrt{2}$, $\mathbf{f}_m^r = (\mathbf{c}_m^r + \mathbf{p}_m^r)/\sqrt{2}$, respectively, where $\mathbf{c}_m^s$, $\mathbf{p}_m^s$, $\mathbf{c}_m^r$, and $\mathbf{p}_m^r$ are the transmit Curtain, transmit Peak, receive reference Curtain, and receive reference Peak sequences, respectively.

Let us consider both periodic and aperiodic AF cases, corresponding to circular and linear time shifts, respectively. For the periodic AF case, the Curtain and Peak sequences satisfy $\mathbf{c}_m^s = \mathbf{c}_m^r = \mathbf{c}_{\xi_m, q_m}$ and $\mathbf{p}_m^s=\mathbf{p}_m^r = \mathbf{p}_m =  \left[p_m[0], p_m[1], \cdots, p_m[N-1]\right]^{\mathrm{T}}$, respectively. For aperiodic AF case, to achieve an ideal curtain structure in the AAF, the sequence length is extended from $N$ to $N+2\bar{\tau}$, where $\bar{\tau} = \tau_{\max}$ denotes the extension length . The Curtain sequences are constructed as $\tilde{\mathbf{c}}_m^s = [\mathbf{0}_{1 \times \bar{\tau}}, \mathbf{c}_{\xi_m, q_m}^T, \mathbf{0}_{1 \times\bar{\tau}}]^T$, $\tilde{\mathbf{c}}_m^r = \overleftrightarrow{\mathbf{c}}_{\xi_m, q_m}$, while the corresponding Peak sequences are obtained by zero-padding as $\tilde{\mathbf{p}}_m^s=\tilde{\mathbf{p}}_m^r=\tilde{\mathbf{p}}_m=\left[\mathbf{0}_{1 \times \bar{\tau }}, \mathbf{p}_m^T, \mathbf{0}_{1 \times \bar{\tau }}\right]^T$. The transmit and receive reference Curtain sequence sets for the periodic and aperiodic cases are defined as $\mathbf{C}^s=\{\mathbf{c}_m^s\}_{m=1}^M$, $\mathbf{C}^r=\{\mathbf{c}_m^r\}_{m=1}^M$, $\tilde{\mathbf{C}}^s=\{\tilde{\mathbf{c}}_m^s\}_{m=1}^M$ and $\tilde{\mathbf{C}}^r=\{\tilde{\mathbf{c}}_m^r\}_{m=1}^M$, respectively. The detailed constructions of the discrete chirp sequences $\mathbf{c}_{\xi_m, q_m}$ and $\overleftrightarrow{\mathbf{c}}\!_{\xi_m, q_m}$, parameterized by the chirp rate $\xi_m$ and phase-shift index $q_m$ are provided in \cite{Flag_sequence}. \looseness=-1\par

We aim to design the Peak sequence set to achieve desirable peak-curtain AF properties of Flag sequences while suppressing interference among different sequences and controlling the peak-to-average power ratio. To this end, we formulate the following optimization problem for the periodic AF case:
\begin{equation}\label{periodic problem}
\begin{array}{ll}
\min\limits_{\mathbf{P}} & G(\mathbf{P}) \\
\text { s.t. } & \left|p_m[n]\right|=1 / \sqrt{N}, \\
& n=0,1, \cdots, N-1, m=1,2, \cdots, M,
\end{array}
\end{equation}
where $\mathbf{P} = \{\mathbf{p}_m\}_{m=1}^M$ is the periodic Peak sequence set, and $G(\mathbf{P})$ jointly measures the masked sidelobe levels of all AAFs and CAFs within $\boldsymbol{\Gamma}$ for the transmit and receive reference sequence sets. Specifically, $G(\mathbf{P})$ is defined as
\begin{equation} \label{WIMSL_objection}
G(\mathbf{P})\!=\!\alpha \!\sum_{m=1}^M\!\! S\left(\mathbf{f}_m^s, \mathbf{f}_m^r\right)\!+\!(1-\alpha)\!\!\!\!\!\sum_{\substack{m_1, m_2 = 1\\
m_2 \neq m_1}}^M \!\!\!\!\!S\!\left(\mathbf{f}_{m_1}^s, \mathbf{f}_{m_2}^r\right),
\end{equation}
where $\alpha \!\in \![0, 1]$ weights the AAFs and CAFs, and
\begin{equation}\label{S expression}
\begin{aligned}
& S\left(\mathbf{f}_{m_1}^s, \mathbf{f}_{m_2}^r\right)=\sum_{(\tau, \omega) \in \mathbf{\Gamma}} \mid W\!\!_{m_1,m_2}(\tau, \omega) \mathbf{p}_{m_1}^{\dagger} \mathbf{U}_{\tau, \omega} \mathbf{p}_{m_2} +\\
&\mathbf{p}_{m_1}^{\dagger}\! \mathbf{U}_{\tau, \omega} \mathbf{c}_{m_2}^r\!+\!\mathbf{c}_{m_1}^s\!\!\!^{\dagger} \mathbf{U}_{\tau, \omega}\mathbf{p}_{m_2}\!+\!\left.\widetilde{W}\!\!_{m_1,m_2} \mathbf{c}_{m_1}^s\!\!\!^{\dagger} \mathbf{U}_{\tau, \omega} \mathbf{c}_{m_2}^r\!\right|^2,
\end{aligned}
\end{equation}
with $\mathbf{U}_{\tau, \omega}=\mathbf{J}_\tau \operatorname{Diag}(\mathbf{h}(\omega))$. Here, $\mathbf{h}(\omega)$ and $\mathbf{J}_{\tau}$ represent the normalized Doppler-shift vector and the circular time-shift matrix\cite{Flag_sequence}, respectively. The matrix $\operatorname{Diag}(\mathbf{h}(\omega))$ is diagonal with $\mathbf{h}(\omega)$ as its diagonal. $W_{m_1,m_2}(\tau,\omega)$ is set to $0$ when $m_1=m_2$ and $(\tau,\omega)=(0,0)$, and to $1$ otherwise. Meanwhile, $\widetilde W_{m_1,m_2}$ equals $0$ when $m_1 = m_2$ and $1$ otherwise. For the periodic AF case, $\mathbf{F}^s = \mathbf{F}^r$ implies $S(\mathbf{f}_{m_1}^s, \mathbf{f}_{m_2}^r) = S(\mathbf{f}_{m_2}^s, \mathbf{f}_{m_1}^r)$. Therefore, only one of the two index pairs $(m_1, m_2)$ and $(m_2, m_1)$, with $m_1 \neq m_2$, needs to be considered in the second summation of \eqref{WIMSL_objection}.
\looseness=-1\par

For the aperiodic AF case, the optimization problem can be formulated as
\begin{equation}\label{aperiodic problem}
\begin{array}{ll}
\min\limits_{\mathbf{\tilde{P}}} & G(\mathbf{\tilde{P}}) \\
\text { s.t. } & \left|{p}_m[n]\right|=1 / \sqrt{N}, \\
& n=0,1, \cdots, N-1, m=1,2, \cdots, M .
\end{array}
\end{equation}
where $\tilde{\mathbf{P}}=\{\tilde{\mathbf{p}}\}_{m=1}^M$ is the aperiodic Peak sequence set, and $G(\tilde{\mathbf{P}})$ is defined analogously to $G(\mathbf{P})$, with the Peak and Curtain sequences replaced by their aperiodic counterparts and the circular time-shift matrix  replaced by the linear time-shift matrix. \looseness=-1\par

Both optimization problems \eqref{periodic problem} and \eqref{aperiodic problem} involve nonconvex objective functions and constraints, making them challenging to solve directly. Moreover, for the aperiodic AF case, the transmit and receive reference Flag sequences are different. Consequently, applying the existing AP-MM algorithm to the aperiodic problem requires relaxation to obtain a formulation similar to that of the periodic AF case. To overcome this limitation, we propose a UM-CG framework that directly addresses the original optimization problems for both cases without relaxation or high-dimensional matrix constructions.
\looseness=-1\par
\vspace{-0.2cm}
\section{Proposed Efficient Solution Algorithm}\label{proposed algorithm}
\vspace{-0.1cm}
In this section, we exploit the geometric structure of the constant-modulus constraints to reformulate the original problems as unconstrained problems on the unimodular manifold. Accordingly, we develop a tailored UM-CG framework based on the Dai-Yuan-type conjugate gradient to solve both problems directly.
\looseness=-1\par

The set of $N \times M$ unimodular matrices forms a unimodular manifold \cite{unimodular}, denoted by $\operatorname{UM}(N, M)$. For any point $\mathbf{X} \in \operatorname{UM}(N, M)$, the tangent space is defined as
\begin{equation}\label{tangent-space}
T_{\mathbf{X}}\operatorname{UM}\left(N,M\right)\!=\!\left\{\!\boldsymbol{\zeta}_{\mathbf{X}}\! \in \!\mathbb{C}^{N \!\times \! M}\!:\!  \operatorname{Re}\left(\boldsymbol{\zeta}_{\mathbf{X}} \!\circ \!\mathbf{X}^* \right)\!=\!\mathbf{0}_{N \times M}\!\right\},
\end{equation}
which characterizes the set of feasible search directions on $\operatorname{UM}(N, M)$ at $\mathbf{X}$. By defining $x_m[n]=\sqrt{N}p_m[n]$, the constant-modulus constraint $|p_m[n]|=1/\sqrt{N}$ is equivalently transformed into $|x_m[n]|=1$. With this variable substitution, the objective function for the periodic AF case, originally formulated in terms of the Peak sequence matrix $\mathbf{P}=[\mathbf{p}_1,\cdots,\mathbf{p}_M]$, can be reformulated in terms of $\mathbf{X}$. For the aperiodic AF case, each Peak sequence $\tilde{\mathbf{p}}_m$ has nonzero entries only at its central $N$ indices. Therefore, the corresponding objective function can be readily expressed in terms of $\mathbf{P}$ and subsequently reformulated in terms of $\mathbf{X}$. The resulting unconstrained optimization problems on $\operatorname{UM}(N,M)$ for the periodic and aperiodic AF cases are respectively expressed as
\begin{equation}\label{Riemannian Optimization Problem}
    \min_{\mathbf{X}\in \operatorname{UM}(N,M)} g(\mathbf{X}), \quad \min_{\mathbf{X}\in \operatorname{UM}(N,M)} g'(\mathbf{X}).
\end{equation}
Here, $g(\mathbf{X})$ and $g'(\mathbf{X})$ are the equivalent reformulations of $G(\mathbf{P})$ and $G(\tilde{\mathbf{P}})$ in terms of $\mathbf{X}$, respectively. To solve the above manifold optimization problems, the Riemannian gradient is required to determine a descent direction on $\operatorname{UM}(N,M)$, while a retraction is employed to map the updated point back onto the manifold. The Riemannian gradient and retraction are respectively defined as
\begin{equation}\label{Riemannian-gradient-operator}
\operatorname{grad}_{\mathbf{X}}g\left( \mathbf{X}\right)\!=\!\nabla_{\mathbf{X}} g\left( \mathbf{X}\right)-\operatorname{Re}\left(\nabla_{\mathbf{X}} g\left(\mathbf{X}\right)\circ \mathbf{X}^*\right)\circ \mathbf{X},
\end{equation}
\begin{equation}\label{retraction-express}
R_{\mathbf{X}}\left(\boldsymbol{\zeta}_{\mathbf{X}}\right)=\left(\mathbf{X}+\boldsymbol{\zeta}_{\mathbf{X}}\right) \circ \frac{1}{\left|\mathbf{X}+\boldsymbol{\zeta}_{\mathbf{X}}\right|_{\left(n,m\right)}},
\end{equation}
where $\boldsymbol{\zeta}_{\mathbf{X}} \in T_{\mathbf{X}}\operatorname{UM}(N, M)$, and the retraction is performed element-wise by normalizing each entry of $(\mathbf{X}+\boldsymbol{\zeta}_{\mathbf{X}})$ to unit modulus. The Euclidean gradient takes the form $\nabla_{\mathbf{X}}g(\mathbf{X}) = \left[\frac{\partial g(\mathbf{X})}{\partial \mathbf{x}_1^*},\cdots,\frac{\partial g(\mathbf{X})}{\partial \mathbf{x}_M^*} \right]$, whose $m$-th column is given by
\begin{equation} \label{Euclidean gradient}
\frac{\partial g(\mathbf{X})}{\partial \mathbf{x}_{m}^*}\! = \! \frac{1}{\sqrt{\!N}}\!\!\left(\!\!\alpha \frac{\partial S(\mathbf{f}_{m}^s, \mathbf{f}_{m}^r)}{\partial \mathbf{p}_{m}^*}\! + \!(1-\alpha)\!\!\!\!\!\!\sum_{\substack{m_1, m_2 = 1\\m_2 \neq m_1}}^M\!\!\!\!\!\frac{\partial S(\mathbf{f}_{m_1}^s,\mathbf{f}_{m_2}^r)}{\partial \mathbf{p}_{m}^*}\!\!\right)
\end{equation}
where
\begin{equation}
\begin{aligned}
\frac{\partial S(\mathbf{f}_{m_1}^s, \mathbf{f}_{m_2}^r)}{\partial \mathbf{p}_{m_1}^*} \!\!=\!\!\!\!\!\! \sum_{(\tau, \omega) \in \boldsymbol{\Gamma}}\!\!\!\!\! A^{*}_{\tau, \omega}(\mathbf{p}_{m_1}, \mathbf{p}_{m_2}) \mathbf{U}_{\tau, \omega}(\mathbf{p}_{m_2} \!+ \!\mathbf{c}_{m_2}^r),
\end{aligned}
\end{equation}
\begin{equation}
\begin{aligned}
\frac{\partial S(\mathbf{f}_{m_1}^s, \mathbf{f}_{m_2}^r)}{\partial \mathbf{p}_{m_2}^*} \!\!=\!\!\!\!\!\! \sum_{(\tau, \omega) \in \boldsymbol{\Gamma}}\!\!\!\!\! A_{\tau, \omega}(\mathbf{p}_{m_1}, \mathbf{p}_{m_2}) \mathbf{U}_{\tau, \omega}^{\dagger}(\mathbf{p}_{m_1} \!+\! \mathbf{c}_{m_1}^s),
\end{aligned}
\end{equation}
\begin{figure*}[t]
 \begin{equation}
 \frac{\partial S(\mathbf{f}_{m}^s, \mathbf{f}_{m}^r)}{\partial \mathbf{p}_{m}^*}\!\! = \!\!\!\!\!\!\sum_{(\tau, \omega) \in \boldsymbol{\Gamma}}\!\!\!\! \left(A^{*}_{\tau, \omega}(\mathbf{p}_{m}, \mathbf{p}_m)\!\left(W_{m, m}(\tau, \omega)\mathbf{U}_{\tau, \omega}\mathbf{p}_m\! +\! \mathbf{U}_{\tau, \omega}\mathbf{c}_m^r\right) \!+\! A_{\tau, \omega}(\mathbf{p}_{m}, \mathbf{p}_m)\!\left(W_{m, m}(\tau, \omega)\mathbf{U}_{\tau, \omega}^{\dagger}\mathbf{p}_m \!+\! \mathbf{U}_{\tau, \omega}^{\dagger}\mathbf{c}_m^s \right)\right),
 \end{equation}
 \begin{equation} \label{Euclidean Part}
 A_{\tau, \omega}(\mathbf{p}_{m_1}, \mathbf{p}_{m_2}) = W_{m_1,m_2}(\tau, \omega) \mathbf{p}_{m_1}\!^{\dagger} \mathbf{U}_{\tau, \omega} \mathbf{p}_{m_2} +\mathbf{p}_{m_1}\!^{\dagger} \mathbf{U}_{\tau, \omega} \mathbf{c}_{m_2}^r+\mathbf{c}_{m_1}^s\!^{\dagger} \mathbf{U}_{\tau, \omega}\mathbf{p}_{m_2}\!+\widetilde{W}_{m_1,m_2} \mathbf{c}_{m_1}^s\!^{\dagger} \mathbf{U}_{\tau, \omega} \mathbf{c}_{m_2}^r.
\end{equation}
\end{figure*}
In the aperiodic setting, the $m$-th column of the Euclidean gradient $\nabla_{\mathbf{X}}g'(\mathbf{X})$ can be obtained by replacing $\mathbf{p}_m$, $\mathbf{c}_m^s$, and $\mathbf{c}_m^r$ in \eqref{Euclidean gradient}--\eqref{Euclidean Part} with $\tilde{\mathbf{p}}_m$, $\tilde{\mathbf{c}}_m^s$, and $\tilde{\mathbf{c}}_m^r$, respectively, and retaining only the central $N$ entries of the resulting derivative. Once the Euclidean gradient $\nabla_{\mathbf{X}}g'(\mathbf{X})$ is obtained, the corresponding Riemannian gradient $\operatorname{grad}_{\mathbf{X}}g'(\mathbf{X})$ follows directly from \eqref{Riemannian-gradient-operator} by replacing $\nabla_{\mathbf{X}}g(\mathbf{X})$ with $\nabla_{\mathbf{X}}g'(\mathbf{X})$.

In general, nonlinear conjugate gradient directions are more effective than the steepest descent direction with comparable complexity\cite{Numerical_Optimization}. For the optimization problems formulated in \eqref{Riemannian Optimization Problem}, we adopt the Riemannian conjugate gradient \cite{R-CG} and develop a UM-CG algorithm based on the Dai-Yuan-type conjugate gradient scheme, whose update rules are formulated as
\begin{equation}
    \mathbf{X}_{k+1} = R_{\mathbf{X_k}}(t_k\boldsymbol{\eta}_k)
\end{equation}
where $t_k$ denotes the step size at the $k$-th iteration. For $k>0$, the search direction $\boldsymbol{\eta}_k$ is computed as
\begin{equation}\label{search direction}
     \boldsymbol{\eta}_k = -\!\operatorname{grad}\!_{\mathbf{X}_k}g(\mathbf{X}_k) \!+\! \beta_k \mathcal{T}_{t_{k-1}\boldsymbol{\eta}_{k-1}}^{(k-1)}(\boldsymbol{\eta}_{k-1}),
\end{equation}
where $\beta_k$ is the Dai-Yuan-type conjugate, yielding
 \begin{equation} \label{conjugate parameter}
     \beta_k =\frac{\|\operatorname{grad}_{\mathbf{X}_k}g(\mathbf{X}_k)\|_{\mathbf{X}_k}^2}{y_k-\langle\operatorname{grad}_{\mathbf{X}_{k-1}}g(\mathbf{X}_{k-1}), \boldsymbol{\eta}_{k-1}\rangle_{\mathbf{X}_{k-1}}}.
\end{equation}
with $y_k = \langle \operatorname{grad}_{\mathbf{X}_k}g(\mathbf{X}_k), \mathcal{T}_{t_{k-1}\boldsymbol{\eta}_{k-1}}^{(k-1)}(\boldsymbol{\eta}_{k-1})\rangle_{\mathbf{X}_k}$.
Since tangent spaces at different points on a Riemannian manifold are generally distinct vector spaces, addition and the inner product between $\boldsymbol{\eta}_{k-1}$ and $\operatorname{grad}_{\mathbf{X}_k}g(\mathbf{X}_k)$ cannot be directly performed when computing $\boldsymbol{\eta}_k$ and $y_k$. Therefore, the vector transport induced by differentiated retraction\cite{Absil} is employed to map $\boldsymbol{\eta}_{k-1}$ onto the tangent space $T_{\mathbf{X}_k}\operatorname{UM}(N, M)$ as
\begin{equation}
\begin{aligned}
   &\mathcal{T}_{t_{k-1}\boldsymbol{\eta}_{k-1}}^{R}(\boldsymbol{\eta}_{k-1}) = \\
   &\left(\boldsymbol{\eta}_{k-1} - \operatorname{Re}(\boldsymbol{\eta}_{k-1}\circ \mathbf{X}_k^*)\circ \mathbf{X}_k\right)\circ \frac{1}{|\mathbf{X}_{k-1}+ t_{k-1}\boldsymbol{\eta}_{k-1}|_{(n, m)}}
   \end{aligned}
\end{equation}
To ensure the convergence of the proposed algorithm, we adopt the scaled vector transport\cite{scaled_vector_transport} at the $k$-th iteration when $\left\|\mathcal{T}_{t_{k-1} \boldsymbol{\eta}_{k-1}}^R\left(\boldsymbol{\eta}_{k-1}\right)\right\|_{\mathbf{X}_{k}} \geq\left\|\boldsymbol{\eta}_{k-1}\right\|_{\mathbf{x}_{k-1}}$, leading to
\begin{equation}\label{scaled vector transport}
    \mathcal{T}_{t_{k-1}\boldsymbol{\eta}_{k-1}}^{0}(\boldsymbol{\eta}_{k-1}) = \frac{\|\boldsymbol{\eta}_{{k-1}}\|_{\mathbf{X}_{k-1}}}{\|\mathcal{T}_{t_{k-1}\boldsymbol{\eta}_{k-1}}^{R}(\boldsymbol{\eta}_{k-1})\|_{\mathbf{X}_k}}\mathcal{T}_{t_{k-1}\boldsymbol{\eta}_{k-1}}^{R}(\boldsymbol{\eta}_{k-1}),
\end{equation}

The UM-CG algorithm framework for the peak sequence optimization problems \eqref{periodic problem} and \eqref{aperiodic problem} is summarized in \textbf{Table}~\ref{UM-CG}. The same procedure applies to problem \eqref{aperiodic problem} upon replacing $\mathbf{P}$, $\mathbf{C}^s$, $\mathbf{C}^r$ and $\operatorname{grad}_{\mathbf{X}_k}g(\mathbf{X}_k)$ in \eqref{search direction} with $\tilde{\mathbf{P}}$, $\tilde{\mathbf{C}}^s$, $\tilde{\mathbf{C}}^r$ and $\operatorname{grad}_{\mathbf{X}_k}g'(\mathbf{X}_k)$, respectively.

\begin{table}[t]
\renewcommand \arraystretch{1.2}
\caption{The proposed UM-CG algorithm framework }
\label{UM-CG}
\centering
\begin{tabular}{l}
 \hline\hline
 {\bf Initialize:} Peak sequence set $\mathbf{P}$, ZoO $\boldsymbol{\Gamma}$, Curtain sequence set $\mathbf{C}^s, \mathbf{C}^r$.\\
\hspace{0.2cm}  Construct $\mathbf{X}_0\in \operatorname{UM}(N, M)$ based on $\mathbf{P}$.\\
 \hspace{0.2cm} Calculate the Riemannian gradient $\operatorname{grad}_{\mathbf{X}_0}g(\mathbf{X}_0)$ via \eqref{Riemannian-gradient-operator}, \eqref{Euclidean gradient}--\eqref{Euclidean Part}.\\
 \hspace{0.2cm} Set $\boldsymbol{\eta}_0 = -\operatorname{grad}_{\mathbf{X}_0}g(\mathbf{X}_0)$ and update $\mathbf{X}_1$ from $\mathbf{X}_0$ as \\
 \hspace{2cm} $\mathbf{X}_1 = R_{\mathbf{X}_0}(t_0\boldsymbol{\eta}_0)$.\\
 {\bf Iterate:} for $k = 1,2,...$ \\
  \hspace{0.2cm} Calculate the Riemannian gradient $\operatorname{grad}_{\mathbf{X}_k}g(\mathbf{X}_k)$  via \eqref{Riemannian-gradient-operator}, \eqref{Euclidean gradient}--\eqref{Euclidean Part}. \\
  \hspace{0.2cm} \bf{if} $\left\|\mathcal{T}_{t_{k-1} \boldsymbol{\eta}_{k-1}}^R\left(\boldsymbol{\eta}_{k-1}\right)\right\|_{\mathbf{X}_{k}} \leq\left\|\boldsymbol{\eta}_{k-1}\right\|_{\mathbf{x}_{k-1}}$\\
  \hspace{2cm} $\mathcal{T}_{t_{k-1} \boldsymbol{\eta}_{k-1}}^{(k-1)}\left(\boldsymbol{\eta}_{k-1}\right)= \mathcal{T}_{t_{k-1} \boldsymbol{\eta}_{k-1}}^R\left(\boldsymbol{\eta}_{k-1}\right),$\\
  \hspace{0.2cm} \bf{else}\\
  \hspace{2cm} $\mathcal{T}_{t_{k-1} \boldsymbol{\eta}_{k-1}}^{(k-1)}\left(\boldsymbol{\eta}_{k-1}\right)= \mathcal{T}_{t_{k-1} \boldsymbol{\eta}_{k-1}}^0\left(\boldsymbol{\eta}_{k-1}\right),$\\
  \hspace{0.2cm} \bf{end if}\\
  \hspace{0.2cm} Calculate the search direction $\boldsymbol{\eta}_k$ via \eqref{conjugate parameter} as\\
  \hspace{2cm}$\boldsymbol{\eta}_k = -\operatorname{grad}_{\mathbf{X}_k}g(\mathbf{X}_k) + \beta_k \mathcal{T}_{t_{k-1}\boldsymbol{\eta}_{k-1}}^{(k-1)}(\boldsymbol{\eta}_{k-1})$\\
  \hspace{0.2cm} Update $\ \mathbf{X}_{k+1}$ from $\mathbf{X}_k$ as .\\
  \hspace{2cm} $\mathbf{X}_{k+1} = R_{\mathbf{X}_k}(t_k\boldsymbol{\eta}_k)$ \\
  {\bf Until} some preset termination criterion is satisfied.\\
  \hspace{0.2cm} \bf{end}\\
  \hspace{0.2cm} Get the Peak sequence set $\mathbf{P}$ from $\mathbf{X}_{k+1}$\\
 \hline\hline
\end{tabular}
\end{table}
\vspace{-0.2cm}
\section{Performance Analysis} \label{performance analysis}
\vspace{-0.1cm}
In this section, we first analyze the computational complexity of the proposed algorithm and the dimensions of the matrices stored during iterations. Furthermore, we theoretically establish its global convergence property under mild conditions.
\vspace{-0.3cm}
\subsection{Implementation Analysis}
\vspace{-0.1cm}
The main computational cost of the proposed UM-CG algorithm arises from evaluating the Euclidean gradient $\nabla_{\mathbf{X}}g(\mathbf{X})$. The computations in \eqref{Euclidean gradient}--\eqref{Euclidean Part} share the same structure as the variable updates in the AP-MM algorithm and can therefore be efficiently implemented using the fast Fourier transform\cite{MM_Song}. Accordingly, the per-iteration complexity of UM-CG is on the order of $\mathcal{O}(M^2KN\log N)$, matching that of a single variable update in the AP-MM algorithm \cite{Flag_sequence}. However, AP-MM requires two variable updates per iteration, each involving a full computation, whereas UM-CG requires only one, resulting in a lower computational cost per iteration.

Furthermore, each AP-MM variable update requires constructing and storing a $2MN\times 2MN$ matrix to determine the parameter for majorizing the objective function from its maximum-magnitude entry, incurring substantial memory overhead. In contrast, UM-CG only requires storing matrices of size at most $N\times M$, thereby significantly reducing memory requirements.

\vspace{-0.3cm}
\subsection{Convergence Analysis}
\vspace{-0.1cm}
\begin{theorem} \label{convergence_analysis}
Let $\{\mathbf{X}_k\}$ be a sequence generated by the proposed
UM-CG algorithm in \textbf{Table} \ref{UM-CG}.
If the step sizes $t_k$ satisfy the following conditions
for $0<c_1<c_2<1$:
\begin{subequations}
    \begin{align}
    & g(\mathbf{X}_{k+1})
    \leq g(\mathbf{X}_k)
    + c_1 t_k
    \langle \operatorname{grad}_{\mathbf{X}_k}\!g(\mathbf{X}_k),
    \boldsymbol{\eta}_k \rangle\!_{\mathbf{X}_k}, \label{stepsize condition 1} \\
    & \langle \operatorname{grad}_{\mathbf{X}_{k+1}}\!g(\mathbf{X}_{k+1}), \!
    \mathcal{T}\!^R_{t_k\boldsymbol{\eta}_k}\!
    (\boldsymbol{\eta}_k)
    \rangle\!_{\mathbf{X}_{k+1}} \!\!\geq\!
    c_2 \langle
    \operatorname{grad}_{\mathbf{X}_k}\!g(\mathbf{X}_k),\!
    \boldsymbol{\eta}_k
    \rangle\!_{\mathbf{X}_k}, \label{stepsize condition 2}
    \end{align}
\end{subequations}
then the algorithm globally converges in the sense that $\liminf_{k\rightarrow\infty}\|\operatorname{grad}_{\mathbf{X}_k} g(\mathbf{X}_k)\|_{\mathbf{X}_k}=0.$
\end{theorem}
\emph{Proof}: This proof is presented in Appendix~\ref{convergence proof}.
\vspace{-0.2cm}
\section{Numerical Results} \label{numerical results}
\vspace{-0.1cm}
In this section, we characterize the overall performance of the proposed UM-CG algorithm. We set $M = 3$ and $N = 1021$, with the maximum delay and Doppler shifts within the ZoO as $\tau_{\max} = 10$ and $\omega_{\max} = 10$, respectively. Three Weil sequences\cite{Flag_sequence} are employed as the initial Peak sequences, while the Curtain sequences are specified by $\xi_1 = q_1 = -1$, $\xi_2 = q_2 = 2$, and $\xi_3 = q_3 = 1$. In what follows, we investigate the convergence behavior of the proposed UM-CG algorithm under both periodic and aperiodic AF cases compared to the AP-MM algorithm. Subsequently, we evaluate the AAF and CAF of the flag sequences designed using the proposed algorithm. \looseness=-1\par

\begin{figure}[!t]
    \centering
     \includegraphics[scale = 0.5]{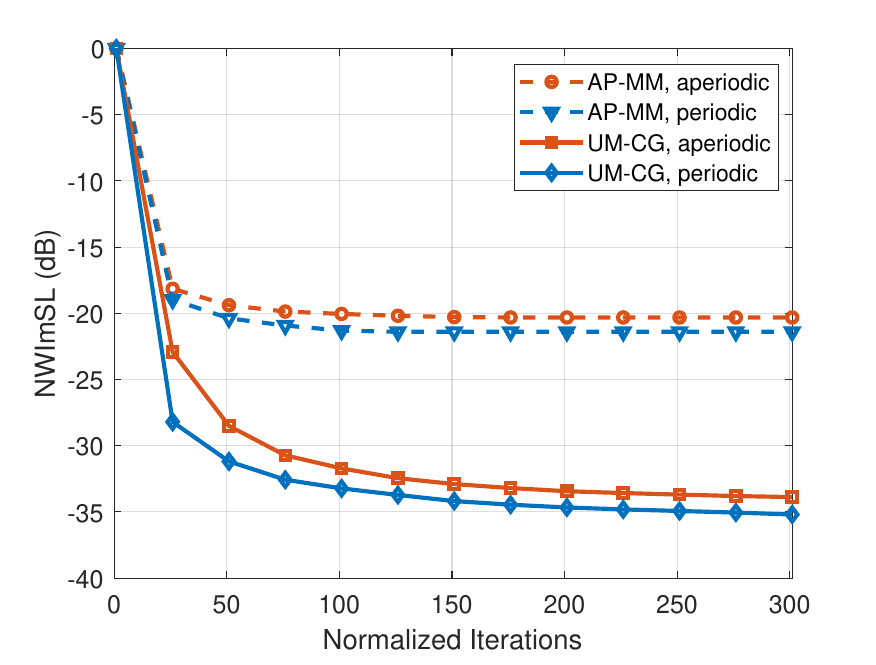}
    \caption{NWImSL Evolution curves of the proposed UM-CG and compared to the conventional AP-MM.}
    \label{NWImSL curve}
\end{figure}
To evaluate the AF performance and characterize the decrease in the objective function during optimization, we define the normalized weighted integrated masked sidelobe level at iteration $k$ for periodic case as $\operatorname{NWImSL}_k = 10\log_{10}(g(\mathbf{X}_k)/g(\mathbf{X_0}))$. For the aperiodic case, $g(\mathbf{X}_k)$ and $g(\mathbf{X}_0)$ are replaced by $g'(\mathbf{X}_k)$ and $g'(\mathbf{X}_0)$, respectively. Fig.~\ref{NWImSL curve} illustrates the convergence behavior of the proposed UM-CG and AP-MM algorithms in terms of NWImSL versus the normalized iteration count. For a fair comparison, the iteration counts are normalized to account for the two variable updates per AP-MM iteration versus one per UM-CG iteration. It can be observed that UM-CG monotonically decreases in NWImSL and gradually converges in both periodic and aperiodic AF cases, achieving lower NWImSL than AP-MM. Moreover, unlike AP-MM, the proposed UM-CG offers a theoretical global convergence guarantee and directly handles aperiodic AF optimization without simplifying the objective function into its periodic formulation. Together with the theoretical analysis provided in \textbf{Theorem \ref{convergence_analysis}}, these results demonstrate the advantages of the proposed UM-CG in terms of optimization performance, convergence guarantees, and applicability to both periodic and aperiodic AF optimization.
\begin{figure*}[htbp]
\centering
\subfigure[]{
\begin{minipage}[t]{0.3\linewidth}
\centering
\centerline{\includegraphics[scale=0.4]{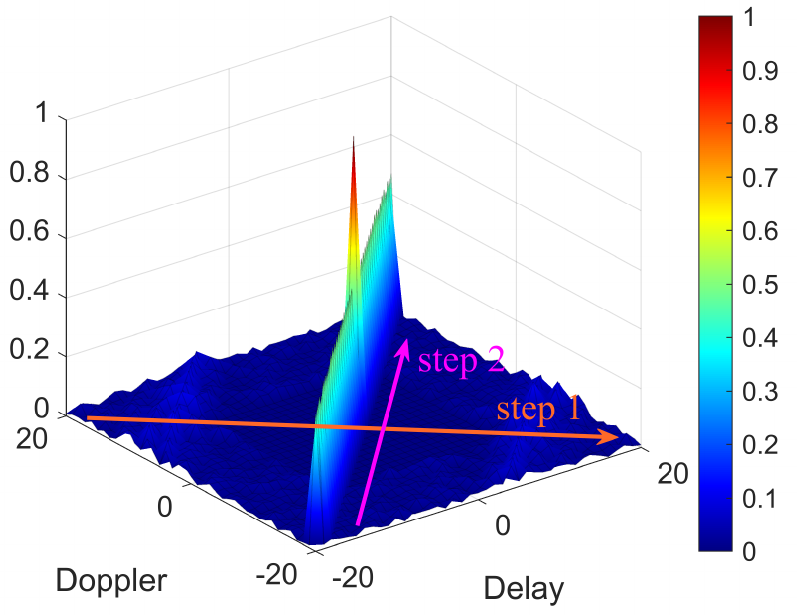}}
\end{minipage}%
}%
\subfigure[]{
\begin{minipage}[t]{0.3\linewidth}
\centering
\centerline{\includegraphics[scale=0.4]{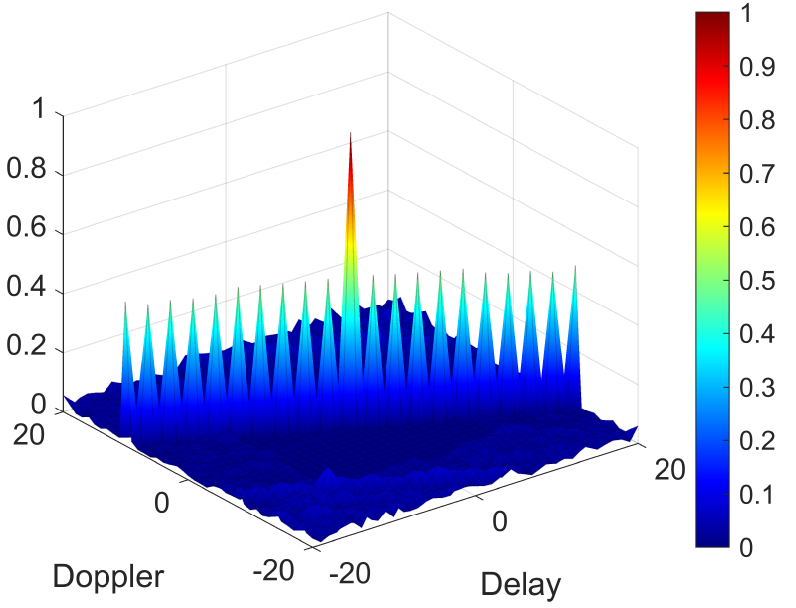}}
\end{minipage}
}
\subfigure[]{
\begin{minipage}[t]{0.3\linewidth}
\centering
\centerline{\includegraphics[scale=0.4]{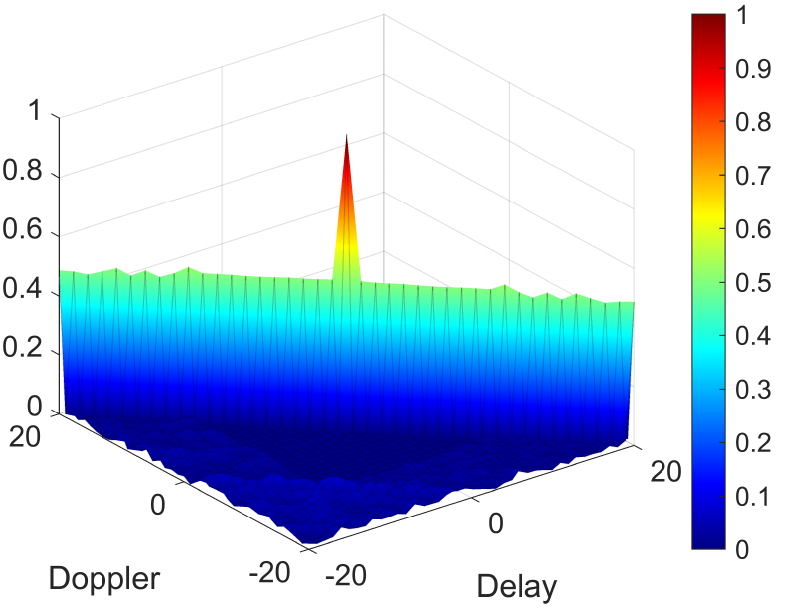}}
\end{minipage}
}
\caption{Periodic AAFs of the Flag sequence set designed by UM-CG: (a) $\mathbf{A}_{\mathbf{f}_1, \mathbf{f}_1}$, (b) $\mathbf{A}_{\mathbf{f}_2, \mathbf{f}_2}$, (c) $\mathbf{A}_{\mathbf{f}_3, \mathbf{f}_3}$. }
\label{AAF figure}
\end{figure*}

\begin{figure*}[htbp]
\centering
\subfigure[]{
\begin{minipage}[t]{0.3\linewidth}
\centering
\centerline{\includegraphics[scale=0.4]{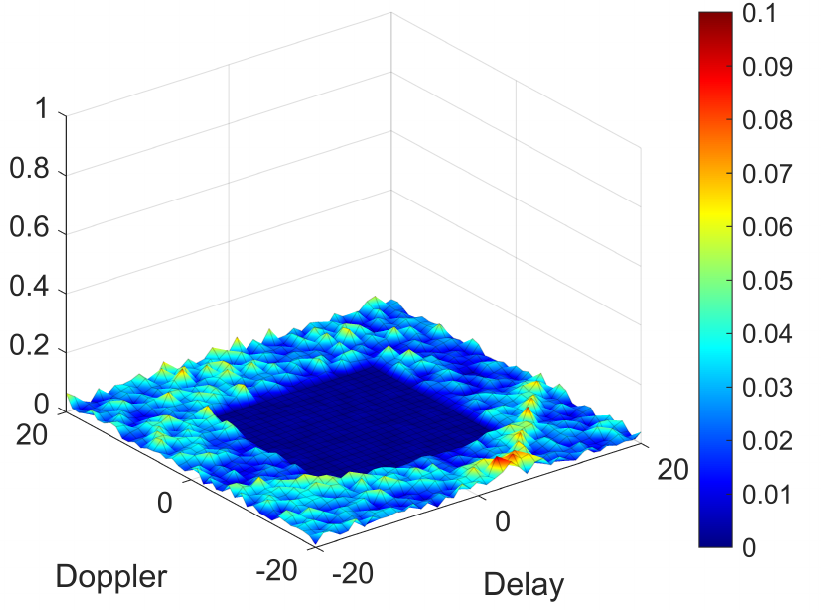}}
\end{minipage}%
}%
\subfigure[]{
\begin{minipage}[t]{0.3\linewidth}
\centering
\centerline{\includegraphics[scale=0.4]{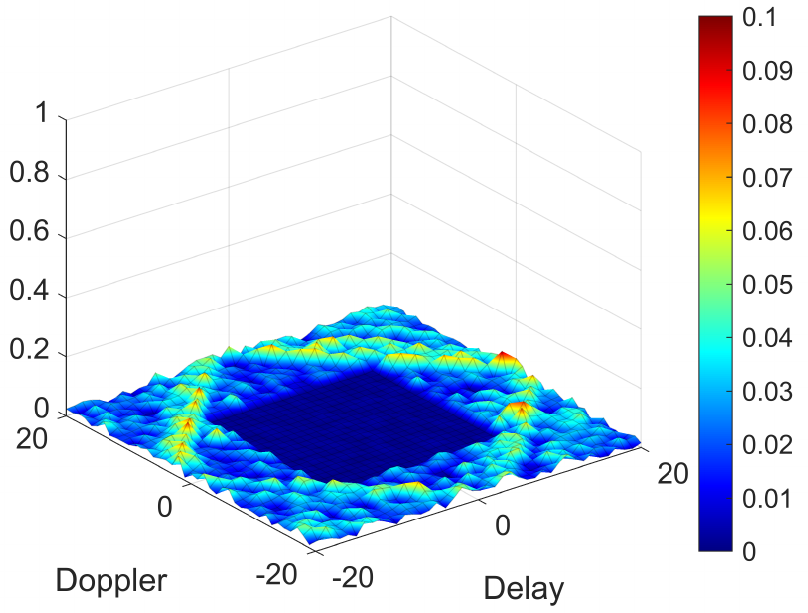}}
\end{minipage}
}
\subfigure[]{
\begin{minipage}[t]{0.3\linewidth}
\centering
\centerline{\includegraphics[scale=0.4]{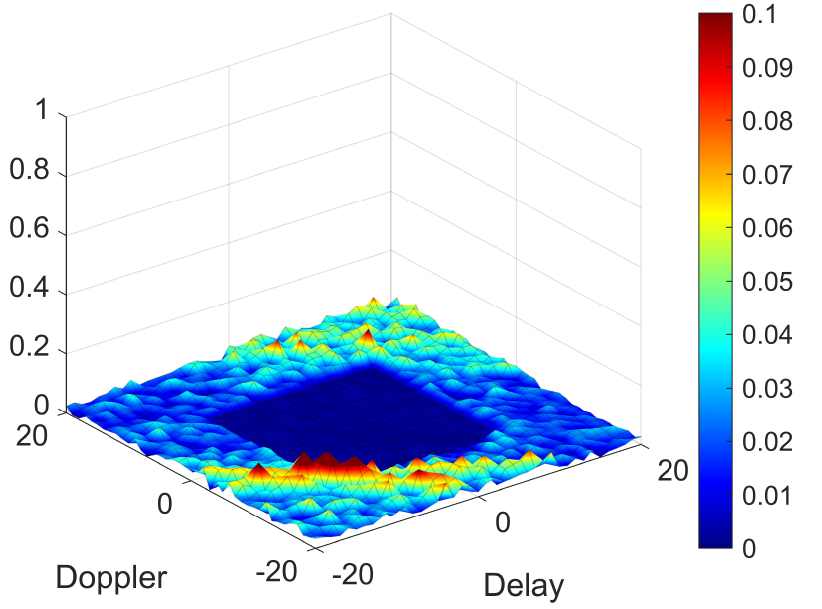}}
\end{minipage}
}
\caption{Periodic CAFs of the Flag sequence set designed by UM-CG: (a) $\mathbf{A}_{\mathbf{f}_1, \mathbf{f}_2}$, (b) $\mathbf{A}_{\mathbf{f}_1, \mathbf{f}_3}$, (c) $\mathbf{A}_{\mathbf{f}_2, \mathbf{f}_3}$. }
\label{CAF figure}
\end{figure*}
Figs.~\ref{AAF figure} and \ref{CAF figure} present the AAFs and CAFs of the Flag sequences designed by the proposed UM-CG, respectively. The $(\tau, \omega)$-th entry of the AAF for the $m$-th Flag sequence is expressed as $\mathbf{A}_{\mathbf{f}_m, \mathbf{f}_m}[\tau, \omega] = |\mathbf{f}_m^{\dagger}\mathbf{U}_{\tau, \omega}\mathbf{f}_m|$. The corresponding entry of the CAF matrix between the $m_1$-th and $m_2$-th Flag sequences is expressed as $\mathbf{A}_{\mathbf{f}_{m_1}, \mathbf{f}_{m_2}}[\tau, \omega] = |\mathbf{f}_{m_1}^{\dagger}\mathbf{U}_{\tau, \omega}\mathbf{f}_{m_2}|$. We observe that the AAFs exhibit a nearly ideal peak-curtain structure with low sidelobe levels within the ZoO, while the CAFs of the Flag sequence set exhibit near-zero sidelobe levels in the same region. Furthermore, as shown in Fig.~\ref{AAF figure}(a), the curtain is first identified in Step 1, followed by peak localization in Step 2, enabling delay-Doppler estimation with nearly linear complexity.

\vspace{-0.3cm}
\section{Conclusion} \label{conclusion}
\vspace{-0.1cm}
We proposed a computationally efficient UM-CG framework for designing Flag sequence sets exhibiting near-ideal peak-curtain ambiguity characteristics. By exploiting the geometry of the unimodular manifold, the proposed algorithm directly minimizes the weighted integrated masked sidelobe level while avoiding excessive computational and storage overhead of the conventional AP-MM algorithm. Furthermore, its global convergence was theoretically established under mild conditions. Numerical results demonstrated the superior optimization performance of UM-CG and confirmed its capability of generating Flag sequences in achieving nearly ideal peak-curtain AAFs.

\appendices
\section{Proof of Theorem~\ref{convergence_analysis}} \label{convergence proof}
We first show that each search direction $\boldsymbol{\eta}_k$ is a descent direction, provided that $\operatorname{grad}_{\mathbf{X}_k}g(\mathbf{X}_k)\neq 0$, and that the sequence of objective function values $\{g(\mathbf{X}_k)\}$ converges. Subsequently, the global convergence of the proposed UM-CG algorithm is established in the sense that $\liminf_{k\rightarrow\infty}\|\operatorname{grad}_{\mathbf{X}_k} g(\mathbf{X}_k)\|_{\mathbf{X}_k}=0.$

For $k = 0$, we have $\boldsymbol{\eta}_0 = -\operatorname{grad}_{\mathbf{X}_0}g(\mathbf{X}_0)$. It follows directly that
\begin{equation}
    \langle\operatorname{grad}_{\mathbf{X}_0}g(\mathbf{X}_0), \boldsymbol{\eta}_0\rangle_{\mathbf{X}_0} = -\|\operatorname{grad}_{\mathbf{X}_0}g(\mathbf{X}_0)\|_{\mathbf{X}_0}^2 \leq 0.
\end{equation}
Next, we use mathematical induction to prove that the search direction $\boldsymbol{\eta}_k$ is a descent direction for all $k \geq 0$, provided that $\operatorname{grad}_{\mathbf{X}_k}g(\mathbf{X}_k) \neq 0$. Suppose that the result holds at iteration $k$, i.e.,
\begin{equation}\label{decrease assumption}
    \langle\operatorname{grad}_{\mathbf{X}_k}g(\mathbf{X}_k), \boldsymbol{\eta}_k\rangle_{\mathbf{X}_k} < 0.
\end{equation}
For iteration $k+1$, it follows that
\begin{equation}\label{k+1 case}
    \begin{aligned}
    &\langle\operatorname{grad}_{\mathbf{X}_{k+1}}g(\mathbf{X}_{k+1}), \boldsymbol{\eta}_{k+1}\rangle_{\mathbf{X}_{k+1}} \!=\! -\|\!\operatorname{grad}_{\mathbf{X}_{k+1}}\!g(\mathbf{X}_{k+1})\|_{\mathbf{X}_{k+1}}^2\\
    &+\beta_{k+1}\langle\operatorname{grad}_{\mathbf{X}_{k+1}}g(\mathbf{X}_{k+1}), \mathcal{T}_{t_k\boldsymbol{\eta}_k}^{(k)}(\boldsymbol{\eta}_k)\rangle_{\mathbf{X}_{k+1}}.
    \end{aligned}
\end{equation}
By substituting \eqref{conjugate parameter} into \eqref{k+1 case} and simplifying the resulting expression, we can obtain
\begin{equation} \label{decrease direction}
    \begin{aligned}
        &\langle\operatorname{grad}_{\mathbf{X}_{k+1}}g(\mathbf{X}_{k+1}), \boldsymbol{\eta}_{k+1}\rangle_{\mathbf{X}_{k+1}}=\\
        &\frac{\|\operatorname{grad}_{\mathbf{X}_{k+1}}\!g(\mathbf{X}_{k+1})\|_{\mathbf{X}_{k+1}}^2\left\langle\operatorname{grad}_{\mathbf{X}_k}g(\mathbf{X}_k), \boldsymbol{\eta}_k\right\rangle_{\mathbf{X}_k}}{\langle\operatorname{grad}_{\mathbf{X}_{k+1}}\!g(\mathbf{X}_{k+1}), \mathcal{T}_{t_k\boldsymbol{\eta}_k}^{(k)}(\boldsymbol{\eta}_k)\rangle_{\mathbf{X}_{k+1}}\!-\!\langle\operatorname{grad}_{\mathbf{X}_k}g(\mathbf{X}_k), \boldsymbol{\eta}_k\rangle_{\mathbf{X}_k}}.
    \end{aligned}
\end{equation}
If $\langle\operatorname{grad}_{\mathbf{X}_{k+1}}g(\mathbf{X}_{k+1}), \mathcal{T}_{t_k\boldsymbol{\eta}_k}^{(k)}(\boldsymbol{\eta}_k)\rangle_{\mathbf{X}_{k+1}} \geq 0$, it follows from \eqref{decrease assumption} that
\begin{equation}
    \langle\operatorname{grad}_{\mathbf{X}_{k+1}}g(\mathbf{X}_{k+1}), \mathcal{T}_{t_k\boldsymbol{\eta}_k}^{(k)}(\boldsymbol{\eta}_k)\rangle_{\mathbf{X}_{k+1}}\!\!>\!  \langle\operatorname{grad}_{\mathbf{X}_k}g(\mathbf{X}_k), \boldsymbol{\eta}_k\rangle_{\mathbf{X}_k}.
\end{equation}
Hence, the denominator of \eqref{decrease direction} is positive, whereas its numerator is negative, which implies that $\langle\operatorname{grad}_{\mathbf{X}_{k+1}}g(\mathbf{X}_{k+1}), \boldsymbol{\eta}_{k+1}\rangle_{\mathbf{X}_{k+1}}<0$, i.e., \eqref{decrease assumption} also holds at iteration $k+1$. Under the condition of $\langle\operatorname{grad}_{\mathbf{X}_{k+1}}g(\mathbf{X}_{k+1}), \mathcal{T}_{t_k\boldsymbol{\eta}_k}^{(k)}(\boldsymbol{\eta}_k)\rangle_{\mathbf{X}_{k+1}} < 0$, we can derive based on \eqref{scaled vector transport} and condition \eqref{stepsize condition 2} of
\begin{equation}
    \begin{aligned} &\langle\operatorname{grad}_{\mathbf{X}_{k+1}}g(\mathbf{X}_{k+1}), \mathcal{T}_{t_k\boldsymbol{\eta}_k}^{(k)}(\boldsymbol{\eta}_k)\rangle_{\mathbf{X}_{k+1}}\\
    &\geq \langle\operatorname{grad}_{\mathbf{X}_{k+1}}g(\mathbf{X}_{k+1}), \mathcal{T}_{t_k\boldsymbol{\eta}_k}^R(\boldsymbol{\eta}_k)\rangle_{\mathbf{X}_{k+1}}\\
    & \geq c_2 \langle
    \operatorname{grad}_{\mathbf{X}_k}g(\!\mathbf{X}_k),
    \boldsymbol{\eta}_k
    \rangle_{\mathbf{X}_k}\\
    &\geq\langle
    \operatorname{grad}_{\mathbf{X}_k}g(\!\mathbf{X}_k),
    \boldsymbol{\eta}_k
    \rangle_{\mathbf{X}_k}.
    \end{aligned}
\end{equation}
Similarly, the numerator of \eqref{decrease direction} is negative, while the denominator is positive. Therefore, $\langle\operatorname{grad}_{\mathbf{X}_{k+1}}g(\mathbf{X}_{k+1}), \boldsymbol{\eta}_{k+1}\rangle_{\mathbf{X}_{k+1}}<0$ holds.

Therefore, we can readily show that, provided that $\operatorname{grad}_{\mathbf{X}_k}g(\mathbf{X}_k) \neq 0$, the following inequality holds for all $k \geq 0$
\begin{equation}
\langle\operatorname{grad}_{\mathbf{X}_k}g(\mathbf{X}_k),
\boldsymbol{\eta}_k\rangle_{\mathbf{X}_k}<0.
\end{equation}
Hence, all search directions $\boldsymbol{\eta}_k$ generated by the proposed UM-CG algorithm are descent direction, and $\beta_k > 0 $ for all $k \geq 1$.
Since each search direction $\boldsymbol{\eta}_k$ is a descent direction, the condition \eqref{stepsize condition 1} guarantees
\begin{equation}
g(\mathbf{X}_{k})- g(\mathbf{X}_{k+1}) \geq -c_1 t_k
    \langle \operatorname{grad}_{\mathbf{X}_k}g(\mathbf{X}_k),
    \boldsymbol{\eta}_k \rangle_{\mathbf{X}_k}\geq0.
\end{equation}
Consequently, the sequence of objective function values $\{g(\mathbf{X}_k)\}$ is nonincreasing. Since $g(\mathbf{X})\geq 0$ is bounded below, the sequence $\{g(\mathbf{X}_k)\}$ converges.

We proceed to establish the global convergence property of the proposed UM-CG algorithm by proving that $ \liminf_{k\rightarrow\infty}
\|\operatorname{grad}_{\mathbf{X}_k}g(\mathbf{X}_k)\|_{\mathbf{X}_k}=0$. To facilitate the proof, we recall the following lemma\cite{Nicolas} for the manifold $\operatorname{UM}(N, M)$.
\begin{lemma}
    Consider a retraction $R$ on $\operatorname{UM}(N,M)$, a compact subset $\mathcal{K} \subseteq \operatorname{UM}(N, M)$ and a continuous, nonnegative function $r: \mathcal{K} \rightarrow \mathbb{R}$. The set
    \begin{equation}
        \boldsymbol{T} = \{(\mathbf{X, \mathbf{S}})\in T\operatorname{UM}(N,M): \mathbf{X}\in\mathcal{K} \ \text{and} \ \|\mathbf{S}\|\leq r(\mathbf{X})\},
    \end{equation}
    is compact in the tangent bundle $T\operatorname{UM}(N, M)$. Let $g: \operatorname{UM}(N,M) \rightarrow \mathbb{R}$ is twice continuously differentiable. There exists a constant $L_1$ satisfies
    \begin{equation}\label{Lipschitz original}
        \|\operatorname{grad}\hat{g}_{\mathbf{X}}(\mathbf{S})- \operatorname{grad}\hat{g}_{\mathbf{X}}(\mathbf{0})\| \leq L_1\|\mathbf{S}\|,
    \end{equation}
\end{lemma}
for all $(\mathbf{X}, \mathbf{S}) \in \boldsymbol{T}$, with $\hat{g}_{\mathbf{X}} = g \circ R_{\mathbf{X}}$. For $\mathbf{S} = t_k\boldsymbol{\eta}_k$, we obtain
\begin{equation}
     \|\operatorname{grad}\hat{g}_{\mathbf{X}_k}(t_k\boldsymbol{\eta}_k)- \operatorname{grad}\hat{g}_{\mathbf{X}_k}(\mathbf{0})\| \leq L_1t_k\|\boldsymbol{\eta}_k\|_{\mathbf{X}_k},
\end{equation}
leading to
\begin{equation}\label{Inequation from Lipschitz}
\begin{aligned}
    &D(g\circ R_{\mathbf{X}_k})(t_k\boldsymbol{\eta}_k)[\boldsymbol{\eta}_k]-D(g\circ R_{\mathbf{X}_k})(\mathbf{0})[\boldsymbol{\eta}_k]\\
    &\leq |D(g\circ R_{\mathbf{X}_k})(t_k\boldsymbol{\eta}_k)[\boldsymbol{\eta}_k]-D(g\circ R_{\mathbf{X}_k})(\mathbf{0})[\boldsymbol{\eta}_k]|\\
    &\leq \|\operatorname{grad}\hat{g}_{\mathbf{X}_k}(t_k\boldsymbol{\eta}_k)- \operatorname{grad}\hat{g}_{\mathbf{X}_k}(\mathbf{0})\| \|\boldsymbol{\eta}_k\|_{\mathbf{X}_k}\\
    &\leq L_1 t_k\|\boldsymbol{\eta}_k\|_{\mathbf{X}_k}^2.
\end{aligned}
\end{equation}
It follows from condition \eqref{stepsize condition 2} that
\begin{equation}
    \begin{aligned}
        &D(g\circ R_{\mathbf{X}_k})(t_k\boldsymbol{\eta}_k)[\boldsymbol{\eta}_k]\\
        &= Dg(R_{\mathbf{X}_k}(t_k\boldsymbol{\eta}_k))[DR_{\mathbf{X}_k}(t_k\boldsymbol{\eta}_k)[\boldsymbol{\eta}_k]]\\
        &= \langle \operatorname{grad}_{\mathbf{X}_{k+1}}g(\mathbf{X}_{k+1}), DR_{\mathbf{X_k}}(t_k\boldsymbol{\eta}_k)[\boldsymbol{\eta}_k]\rangle_{\mathbf{X}_{k+1}}\\
        &=\langle \operatorname{grad}_{\mathbf{X}\!_{k+1}}g(\mathbf{X}_{k+1}),
    \mathcal{T}^R_{t_k\boldsymbol{\eta}_k}
    (\boldsymbol{\eta}_k)
    \rangle_{\mathbf{X}_{k+1}}\\
    &\geq c_2\langle\operatorname{grad}_{\mathbf{X}_k}g(\mathbf{X}_k), \boldsymbol{\eta}_k \rangle_{\mathbf{X}_k},
    \end{aligned}
\end{equation}
which further implies that
\begin{equation}
    \begin{aligned}
        &D(g\circ R_{\mathbf{X}_k})(t_k\boldsymbol{\eta}_k)[\boldsymbol{\eta}_k]-D(g\circ R_{\mathbf{X}_k})(\mathbf{0})[\boldsymbol{\eta}_k]\\
        &\geq (c_2-1)\langle\operatorname{grad}_{\mathbf{X}_k}g(\mathbf{X}_k), \boldsymbol{\eta}_k \rangle_{\mathbf{X}_k}.
    \end{aligned}
\end{equation}
Combining the inequality \eqref{Inequation from Lipschitz} gives
\begin{equation}\label{stepsize Inequation}
\begin{aligned}
  &L_1t_k\|\boldsymbol{\eta}_k\|^2 \geq (c_2-1)\langle\operatorname{grad}_{\mathbf{X}_k}g(\mathbf{X}_k), \boldsymbol{\eta}_k \rangle_{\mathbf{X}_k}\\
  &t_k \geq \frac{(c_2-1)\langle\operatorname{grad}_{\mathbf{X}_k}g(\mathbf{X}_k), \boldsymbol{\eta}_k \rangle_{\mathbf{X}_k}}{L_1\|\boldsymbol{\eta}_k\|_{\mathbf{X}_k}^2}
\end{aligned}
\end{equation}
Substitution of \eqref{stepsize Inequation} into \eqref{stepsize condition 1} yields
\begin{equation}\label{Zoutendijk}
 g(\mathbf{X}_{k+1})\leq g(\mathbf{X}_k)+ \frac{c_1(c_2-1)}{L_1} \frac{ \langle \operatorname{grad}_{\mathbf{X}_k}g(\mathbf{X}_k),
    \boldsymbol{\eta}_k \rangle_{\mathbf{X}_k}^2}{\|\boldsymbol{\eta}_k\|_{\mathbf{X}_k}^2}
\end{equation}
Upon summing \eqref{Zoutendijk} over all $k$ and invoking the convergence of the objective function sequence $\{g(\mathbf{X}_k)\}$, we conclude that
\begin{equation}\label{Zoutendijk Conclusion}
    \sum_{k = 0}^{\infty}\frac{ \langle \operatorname{grad}_{\mathbf{X}_k}g(\mathbf{X}_k),
    \boldsymbol{\eta}_k \rangle_{\mathbf{X}_k}^2}{\|\boldsymbol{\eta}_k\|_{\mathbf{X}_k}^2} \leq \infty
\end{equation}
Furthermore, \eqref{search direction} implies that
\begin{equation}
\begin{aligned}
    &\|\boldsymbol{\eta}_{k+1}\|_{\mathbf{X}_{k+1}}^2 = \beta_{k+1}^2\|\mathcal{T}_{t_{k}\boldsymbol{\eta}_{k}}^{(k)}(\boldsymbol{\eta}_{k})\|_{\mathbf{X}_{k+1}}^2 - \\
    &2\langle \operatorname{grad}_{\mathbf{X}_{k+1}}g(\mathbf{X}_{k+1}), \boldsymbol{\eta}_{k+1}\rangle_{\mathbf{X}_{k+1}}\!-\!\|\operatorname{grad}_{\mathbf{X}_{k+1}}g(\mathbf{X}_{k+1})\|_{\mathbf{X}_{k+1}}^2
\end{aligned}
\end{equation}
\begin{figure*}
\begin{equation} \label{Inequation from search direction}
\begin{aligned}
&\frac{
\|\boldsymbol{\eta}_{k+1}\|_{\mathbf{X}_{k+1}}^2
}{
\langle
\operatorname{grad}_{\mathbf{X}_{k+1}}g(\mathbf{X}_{k+1}),
\boldsymbol{\eta}_{k+1}
\rangle_{\mathbf{X}_{k+1}}^2
}
=\frac{
\|\mathcal{T}_{t_k\boldsymbol{\eta}_k}^{(k)}
(\boldsymbol{\eta}_k)\|_{\mathbf{X}_{k+1}}^2
}{
\langle
\operatorname{grad}_{\mathbf{X}_k}g(\mathbf{X}_k),
\boldsymbol{\eta}_k
\rangle_{\mathbf{X}_k}^2
}-\frac{2}{
\langle
\operatorname{grad}_{\mathbf{X}_{k+1}}g(\mathbf{X}_{k+1}),
\boldsymbol{\eta}_{k+1}
\rangle_{\mathbf{X}_{k+1}}
}-\frac{
\|\operatorname{grad}_{\mathbf{X}_{k+1}}
g(\mathbf{X}_{k+1})\|_{\mathbf{X}_{k+1}}^2
}{
\langle
\operatorname{grad}_{\mathbf{X}_{k+1}}g(\mathbf{X}_{k+1}),
\boldsymbol{\eta}_{k+1}
\rangle_{\mathbf{X}_{k+1}}^2
}\\
&=\frac{
\|\mathcal{T}_{t_k\boldsymbol{\eta}_k}^{(k)}
(\boldsymbol{\eta}_k)\|_{\mathbf{X}_{k+1}}^2
}{
\langle
\operatorname{grad}_{\mathbf{X}_k}g(\mathbf{X}_k),
\boldsymbol{\eta}_k
\rangle_{\mathbf{X}_k}^2
}-\left(
\frac{1}{
\|\operatorname{grad}_{\mathbf{X}_{k+1}}
g(\mathbf{X}_{k+1})\|_{\mathbf{X}_{k+1}}
}
+\frac{
\|\operatorname{grad}_{\mathbf{X}_{k+1}}
g(\mathbf{X}_{k+1})\|_{\mathbf{X}_{k+1}}
}{
\langle
\operatorname{grad}_{\mathbf{X}_{k+1}}g(\mathbf{X}_{k+1}),
\boldsymbol{\eta}_{k+1}
\rangle_{\mathbf{X}_{k+1}}
}
\right)^2+\frac{1}{
\|\operatorname{grad}_{\mathbf{X}_{k+1}}
g(\mathbf{X}_{k+1})\|_{\mathbf{X}_{k+1}}^2
}\\
&\leq
\frac{
\|\boldsymbol{\eta}_k\|_{\mathbf{X}_k}^2
}{
\langle
\operatorname{grad}_{\mathbf{X}_k}g(\mathbf{X}_k),
\boldsymbol{\eta}_k
\rangle_{\mathbf{X}_k}^2
}
+\frac{1}{
\|\operatorname{grad}_{\mathbf{X}_{k+1}}
g(\mathbf{X}_{k+1})\|_{\mathbf{X}_{k+1}}^2
}.
\end{aligned}
\end{equation}
\end{figure*}
Summation of both sides of \eqref{Inequation from search direction} over $k$ gives the following relation:
\begin{equation}
\begin{aligned}
&\frac{
\|\boldsymbol{\eta}_k\|_{\mathbf{X}_k}^{2}
}{
\left\langle
\operatorname{grad}_{\mathbf{X}_k}g(\mathbf{X}_k),
\boldsymbol{\eta}_k
\right\rangle_{\mathbf{X}_k}^{2}
}\\
&\leq
\sum_{i=1}^{k}
\frac{1}{
\|\operatorname{grad}_{\mathbf{X}_i}
g(\mathbf{X}_i)\|_{\mathbf{X}_i}^{2}
}
+
\frac{
\|\boldsymbol{\eta}_0\|_{\mathbf{X}_0}^{2}
}{
\left\langle
\operatorname{grad}_{\mathbf{X}_0}g(\mathbf{X}_0),
\boldsymbol{\eta}_0
\right\rangle_{\mathbf{X}_0}^{2}
}\\
&\leq
\sum_{i=0}^{k}
\frac{1}{
\|\operatorname{grad}_{\mathbf{X}_i}
g(\mathbf{X}_i)\|_{\mathbf{X}_i}^{2}
}.
\end{aligned}
\end{equation}
If $\liminf_{k\rightarrow\infty}\|\operatorname{grad}_{\mathbf{X}_k}g(\mathbf{X}_k)\|_{\mathbf{X}_k} \neq 0$ for all $k \geq 0$, then there exist a constant $C_1>0$ such that
\begin{equation}
    \|\operatorname{grad_{\mathbf{X}_k}}g(\mathbf{X}_k)\|_{\mathbf{X}_k} \geq C_1
\end{equation}
It can thus be concluded that
\begin{equation}
    \frac{
\|\boldsymbol{\eta}_k\|_{\mathbf{X}_k}^{2}
}{
\left\langle
\operatorname{grad}_{\mathbf{X}_k}g(\mathbf{X}_k),
\boldsymbol{\eta}_k
\right\rangle_{\mathbf{X}_k}^{2}
} \leq \frac{k+1}{C_1}
\end{equation}
\begin{equation}
    \sum_{k = 0}^{\infty}\frac{ \langle \operatorname{grad}_{\mathbf{X}_k}g(\mathbf{X}_k),
    \boldsymbol{\eta}_k \rangle_{\mathbf{X}_k}^2}{\|\boldsymbol{\eta}_k\|_{\mathbf{X}_k}^2} \geq C_1^2 \sum_{k=0}^{\infty}\frac{1}{k+1} = \infty
\end{equation}
which contradicts \eqref{Zoutendijk Conclusion}. Therefore, we have $\liminf_{k\rightarrow\infty}\|\operatorname{grad}_{\mathbf{X}_k} g(\mathbf{X}_k)\|_{\mathbf{X}_k}=0$.

\vfill


\begin{thebibliography}{1}
\bibliographystyle{IEEEtran}
\bibitem{Sui_ISAC}
Z. Sui et al., ``Multi-functional chirp signalling for next-generation multi-carrier wireless networks: Communications, sensing and ISAC perspectives," 2025, \emph{arXiv:2508.06022}.
\bibitem{OTFS_Survey}
H. S. Rou, V. Savaux, Z. Sui, G. T. F. d. Abreu and Z. Liu, ``AFDM: Evolving OFDM Toward 6G+," \emph{IEEE Open Journal of the Communications Society}, vol. 7, pp. 12408-12434, 2026.
\bibitem{Radar_Signal}
M. A. Richards, \emph{Fundamentals of Radar Signal Processing}. New York, NY, USA: McGraw-Hill, 2005.
\bibitem{ZAZ_LAZ}
Z. Ye, Z. Zhou, P. Fan, Z. Liu, X. Lei and X. Tang, ``Low Ambiguity Zone: Theoretical Bounds and Doppler-Resilient Sequence Design in Integrated Sensing and Communication Systems," \emph{IEEE Journal on Selected Areas in Communications}, vol. 40, no. 6, pp. 1809-1822, Jun. 2022.
\bibitem{ideal_AAF}
J. J. Benedetto, I. Konstantinidis, and M. Rangaswamy, ``Phase-coded waveforms and their design," \emph{IEEE Signal Processing Magazine}, vol. 26, no. 1, pp. 22-31, Jan. 2009.
\bibitem{ZoO}
Z. Gu, Z. Zhou, P. Fan, A. R. Adhikary and Z. Liu, ``Oversampled Low Ambiguity Zone Sequences for Channel Estimation Over Doubly Selective Channels,"  \emph{IEEE Transactions on Communications}, vol. 73, no. 4, pp. 2616-2630, Apr. 2025.
\bibitem{CAF}
B. Shen, T. Yu, Z. Zhou, Y. Yang and Z. Liu, ``New Design of Sparse Zero-Correlation-Zone Sequence Sets for Optimal Channel Estimation in (Generalized) Spatial Modulation Systems," \emph{IEEE Transactions on Communications}, vol. 74, pp. 3423-3436, 2026.
\bibitem{delay_doppler_estimation_fast}
J. M. O'Toole, M. Mesbah, and B. Boashash, ``Accurate and efficient implementation of the time-frequency matched filter," \emph{IET Signal Process}., vol. 4, no. 4, pp. 428-437, Aug. 2010.
\bibitem{Linear_complexity}
A. Fish, S. Gurevich, R. Hadani, A. M. Sayeed and O. Schwartz, ``Delay-Doppler Channel Estimation in Almost Linear Complexity," \emph{IEEE Transactions on Information Theory}, vol. 59, no. 11, pp. 7632-7644, Nov. 2013.
\bibitem{Flag_sequence}
L. Meng, Y. L. Guan, Y. Ge and Z. Liu, ``Flag Sequence Set Design for Low-Complexity Delay-Doppler Estimation," \emph{IEEE Transactions on Vehicular Technology}, vol. 74, no. 8, pp. 12131-12147, Aug. 2025.
\bibitem{MM_Song}
J. Song, P. Babu and D. P. Palomar, ``Sequence Design to Minimize the Weighted Integrated and Peak Sidelobe Levels," \emph{IEEE Transactions on Signal Processing}, vol. 64, no. 8, pp. 2051-2064, Apr. 2016.
\bibitem{unimodular}
X. Zhao, J. Wang, S. Yan and Y. Wang, ``Low-Complexity Unimodular Waveform Design With Good Space-Time Correlation for Multitarget Detection in MIMO Radar," \emph{IEEE Transactions on Aerospace and Electronic Systems}, vol. 62, pp. 10834-10850, 2026.
\bibitem{Numerical_Optimization}
J. Nocedal and S. Wright, \emph{Numerical Optimization}. Berlin, Germany: Springer-Verlag, 2006.
\bibitem{R-CG}
H. Sato, ``Riemannian conjugate gradient methods: General framework and specific algorithms with convergence analyses," \emph{SIAM Journal of Optimization}, vol. 32, no. 4, pp. 2690-2717, 2022.
\bibitem{Absil}
P.-A. Absil, R. Mahony and R. Sepulchre, \emph{Optimization Algorithms on Matrix Manifolds}, Princeton, NJ, USA:Princeton Univ. Press, 2009.
\bibitem{scaled_vector_transport}
H. Sato and T. Iwai, ``A new, globally convergent Riemannian conjugate gradient method," \emph{Optimization}, vol. 64, no. 4, pp. 1011-1031, 2015.
\bibitem{Nicolas}
N. Boumal, \emph{An introduction to optimization on smooth manifolds}. Cambridge, U.K.: Cambridge Univ. Press, 2023.
\end{thebibliography}
\end{document}